\documentclass{kluwer}
\usepackage{epsfig,url,graphicx}

\newdisplay{guess}{Conjecture}

\begin{document}
\begin{article}
\begin{opening}
\title{Hybrid XML Retrieval: Combining Information Retrieval and a Native XML Database}
\author{Jovan \surname{Pehcevski} \email{jovanp@cs.rmit.edu.au}}
\institute{RMIT University, Melbourne, Australia}
\author{James A. \surname{Thom} \email{jat@cs.rmit.edu.au}}
\institute{RMIT University, Melbourne, Australia}
\author{Anne-Marie \surname{Vercoustre} \email{anne-marie.vercoustre@inria.fr}}
\institute{INRIA, Rocquencourt, France}

\runningauthor{J. Pehcevski, J. A. Thom and A-M Vercoustre}
\runningtitle{Hybrid XML Retrieval}

\begin{abstract}
This paper investigates the impact of three approaches to XML retrieval:
using Zettair, a full-text information retrieval system; using eXist, a native XML database; and using a hybrid system
that takes full article answers from Zettair and uses eXist to extract elements from those articles.
For the content-only topics, we undertake a preliminary analysis of the INEX 2003 relevance assessments 
in order to identify the types of highly relevant document components.
Further analysis identifies two complementary sub-cases of relevance assessments (\emph{General} and \emph{Specific})
and two categories of topics (\emph{Broad} and \emph{Narrow}).
We develop a novel retrieval module that for a content-only topic
utilises the information from the resulting answer list of a native XML database and
dynamically determines the preferable units of retrieval, which we call \emph{Coherent Retrieval Elements}.
The results of our experiments show that --
when each of the three systems is evaluated against different retrieval scenarios (such as 
different cases of relevance assessments, different topic categories and different choices of evaluation metrics) --
the XML retrieval systems exhibit varying behaviour and 
the best performance can be reached for different values of the retrieval parameters. 
In the case of INEX 2003 relevance assessments for the content-only topics,
our newly developed hybrid XML retrieval system 
is substantially more effective than either Zettair or eXist,
and yields a robust and a very effective XML retrieval.
\end{abstract}

\keywords{XML Information Retrieval, XML Databases, eXist, Zettair, INEX}

\end{opening}

\section{Introduction}

XML retrieval approaches can be classified into three
categories~\cite{INEX02_overview}: approaches that extend well-known full-text \emph{information retrieval} (IR) systems to handle XML retrieval;
 approaches that extend \emph{database
management systems} to deal with XML data; and
\emph{XML-specific} approaches that incorporate XML standards, such as XPath~\cite{XPath}, 
XSL~\cite{XSL} or XQuery~\cite{XQuery}, to handle both XML and full-text search.
For the purpose of exploring the nature of XML-IR,
we propose in this paper a hybrid approach to XML retrieval that combines text information retrieval features
with XML-specific features found in a native XML database.

In this paper we focus on content-only (CO) XML retrieval topics
as part of INEX\footnote{http://www.is.informatik.uni-duisburg.de/projects/inex/index.html.en}, the INitiative for the Evaluation of XML Retrieval.
These topics 
do not explicitly refer to the underlying document structure
and are usually represented by queries constituting a combination of words and phrases.
The major retrieval challenge for the CO topics is
identifying document components considered \emph{relevant} to a user information need
and determining their appropriate level of answer \emph{granularity}
(which corresponds to 
the types of XML document components that are to be returned as answers, or to the \emph{preferable units of retrieval}).

In order to understand which types of document components users consider to be most useful as answers, 
we analyse in Section 2 the INEX 2003 relevance assessments for the CO topics.
In the context of INEX, a full article represents a whole document, while an XML element within an article represents a document component.
Arising from our analysis we identify many cases where, for a particular CO topic and an article,
several layers of elements in the document hierarchy (such as \verb+<article>+, \verb+<bdy>+, \verb+<sec>+ and \verb+<p>+)
have all been assessed as highly relevant, an issue previously referred to as ``overpopulated and varying recall base''~\cite{Overlap}.
This contradicts the structured document retrieval principle outlined in the
FERMI multimedia retrieval model; the principle states that ``the retrieval process, in the context of large amounts
of structured information, has to focus on the smallest units ... that fulfill the query''~\cite{FERMI}.
We therefore investigate two complementary cases of modified relevance assessments: 
one when the relevance assessments only consider the 
\emph{most specific} highly relevant elements within an article (\emph{Specific} relevance assessments), and 
another when the relevance assessments only consider the \emph{most general} highly relevant elements within an article 
(\emph{General} relevance assessments).
In the absence of more realistic user models for XML retrieval, the former case is a close approximation of the FERMI retrieval principle 
and reflects users that prefer specific, more focused answers for their queries, 
whereas the latter case models users that prefer compound and more informative answers for their queries.

An analysis of the CO topics using the General relevance assessments
identifies two categories of topics: topics
where the assessor prefers full articles over more specific elements (\emph{Broad} topics), and
topics where the assessor prefers more specific elements within articles over full articles (\emph{Narrow} topics).

Different retrieval scenarios (using General or Specific relevance assessments and Broad or Narrow topics)
are likely to impact the relative retrieval effectiveness of different XML retrieval systems. 
In Section 3 we consider three systems, based on a full-text information retrieval approach, a native XML database approach,
and a hybrid approach, respectively. 

In the General relevance assessment case, we investigate whether a full-text information retrieval system
is capable of retrieving full articles as highly relevant answers.
Our choice for a full-text information retrieval system is Zettair\footnote{http://www.seg.rmit.edu.au/zettair/}
(formerly know as Lucy),
a compact and fast full-text search engine designed and written by the Search
Engine Group at RMIT University.
Zettair implements an inverted index structure, a well-researched search
structure implemented in many existing full-text information retrieval
systems~\cite{MG99}.
However, in its current implementation Zettair's primary unit of retrieval is a full article, and is not capable of 
indexing and retrieving more specific elements within articles.

In the Specific relevance assessment case, we investigate whether an XML-specific retrieval system
is capable of retrieving the most specific elements as highly relevant answers.
Our choice for an XML-specific retrieval system is eXist\footnote{http://exist-db.org/},
an open-source native XML database.
eXist incorporates most of the basic as well as advanced native XML database
features, such as full and partial keyword text searches, search patterns based on
regular expressions, and keyword proximity functions.
Two notable features are efficient
index-based query processing and XPath extensions for full-text search~\cite{eXist}.
However, eXist, like most native XML database implementations, does not consider ranking of the elements in the resulting answer list.

In order to combine the best retrieval features from the above systems, we propose
a \emph{hybrid} XML retrieval system.
We observe some issues in the native XML database component of our hybrid system that
have a strong impact on its retrieval effectiveness.
We therefore develop and implement a novel retrieval module that, for a CO topic, utilises the information contained in the answer list
of a native XML database
and identifies, ranks and retrieves \emph{Coherent Retrieval Elements (CRE)}.
We regard the Coherent Retrieval elements as preferable units of retrieval for INEX CO topics,
since the results of our experiments show that the CRE module is capable of retrieving
both highly relevant articles and elements within articles at an appropriate level of retrieval granularity.

Our results in Section 4 show that, given the various possible combinations of XML retrieval scenarios
(such as different cases of assessments, topic categories and choices of evaluation metrics),
the XML retrieval systems subject to our analysis exhibit varying retrieval behaviour.
Although further tuning of individual system parameters may improve retrieval performance,
we also show that, in the case of INEX 2003 relevance assessments for the CO topics,
our newly developed Hybrid-CRE system yields robust and effective content-oriented XML retrieval.

Some existing XML retrieval systems have similarities to the approaches presented in this paper. 
We review these systems in Section 5, where we also provide a comparison to other analysis of the INEX CO topics and relevance assessments.

\section{Analysis of INEX 2003 CO Topics and Relevance Assessments}

INEX provides
a means to evaluate the effectiveness of different XML retrieval systems.
The XML document collection used in INEX comprises 12107 IEEE Computer Society
articles published in the period between 1997-2002 with approximately 500MB of data.
Each year (starting in 2002) a new set of ad-hoc XML retrieval topics are introduced
and assessed by the participants.
In 2003, in the light of the experience of the 2002 INEX workshop,
revised relevance dimensions, \emph{exhaustivity} (how many aspects of the topic are covered in the element) and \emph{specificity}
(how specific to the topic is the element),
were introduced to assess the relevance of articles and elements within articles to these topics.

Two types of XML retrieval topics are explored in
INEX: content-only (CO) topics and content-and-structure (CAS) topics.
A CAS topic enforces restrictions with respect to the
existing document structure and explicitly specifies the type of the unit of
retrieval (section, paragraph, or other), whereas a CO topic has no such restriction
on the elements retrieved.
In this paper we focus on improving XML retrieval for CO topics.

\subsection{An INEX CO topic example}

\renewcommand\baselinestretch{}

\begin{table}[tb]
\begin{tabular}{l}
\hline
$<?$xml\ version=``1.0''\ encoding=``ISO-8859-1''$?>$ \\
$<!$DOCTYPE\ inex\_topic\ SYSTEM\ ``topic.dtd''$>$ \\
$<$inex\_topic\ topic\_id=``117''\ query\_type=``CO''\ ct\_no=``98''$>$ \\
\ \ \ $<$title$>$ \\
\ \ \ \ \ \ \ \ Patricia Tries \\
\ \ \ $<$/title$>$ \\
\ \ \ $<$description$>$ \\
\ \ \ \ \ \ \ \ Find documents/elements that describe Patricia tries and their use. \\
\ \ \ $<$/description$>$ \\
\ \ \ $<$narrative$>$ \\
\ \ \ \ \ \ \ \ To be relevant, a document/element must deal with the use of Patricia \\
\ \ \ \ \ \ \ \ Tries for text search. Description of the standard algorithm, \\
\ \ \ \ \ \ \ \ optimised implementation and use in Information retrieval applications \\
\ \ \ \ \ \ \ \ are all relevant. \\
\ \ \ $<$/narrative$>$ \\
\ \ \ $<$keywords$>$ \\
\ \ \ \ \ \ \ \ Patricia tries, tries, text search, string search algorithm, \\
\ \ \ \ \ \ \ \ string pattern matching \\
\ \ \ $<$/keywords$>$ \\
$<$/inex\_topic$>$ \\
\hline
\end{tabular}
\caption{INEX 2003 CO Topic 117}
\label{fig-INEX117}
\end{table}

\renewcommand\baselinestretch{1.5}

Table~\ref{fig-INEX117} shows INEX topic 117, which was proposed by our group.
This topic calls for articles or elements within articles focusing on algorithms
that use Patricia tries for text search.
As indicated in the topic narrative, an article or an element within article is
considered relevant if it describes the standard algorithm,
its optimised implementation, or discusses its usage in information retrieval applications.

This particular topic raises some interesting issues for retrieval, such as:
\begin{itemize}
\item \emph{Patricia} (usually) represents a person's first name, rather than a
data structure;
\item \emph{tries} is a verbal form; and
\item keywords like \emph{text}, \emph{string}, and \emph{search} appear almost
everywhere in the INEX XML document collection.
\end{itemize}

\subsection{CO topics relevance assessments analysis}

The idea behind our analysis of INEX 2003 CO relevance assessments stems from
the challenge of XML content-only retrieval:
since a CO topic does not restrict the answer elements,
the final answer list may contain elements of different types
with varying sizes and granularities.
We observed that full articles may represent preferable answers for some topics,
while for other topics more specific elements within articles may be preferable over full articles.
Whether smaller (more specific) or larger (more general) elements are preferred depends, to some
extent, on the inclinations of the author-come-assessor of a particular topic as to which elements
constitute the best answers to the information need.
This may be apparent from the wording of the topic, especially the narrative.

By analysing the relevance assessments for the CO topics, we aim to understand
what users (that is, the topic authors who later assess the returned answer elements)
consider to be the most useful.

\renewcommand\baselinestretch{}

\begin{table}[tb]
\begin{tabular}{l}
\hline
$<$file\ file=``ic/1999/w4095''$>$ \\
\ \ \ \ \ \ \ $<$path\ E=``3''\ S=``3''\ path=``/article[1]''/$>$ \\
\ \ \ \ \ \ \ $<$path\ E=``3''\ S=``3''\ path=``/article[1]/bdy[1]''/$>$ \\
\ \ \ \ \ \ \ . . . . . . . \\
\ \ \ \ \ \ \ $<$path\ E=``3''\ S=``3''\ path=``/article[1]/bdy[1]/sec[2]''/$>$ \\
\ \ \ \ \ \ \ $<$path\ E=``3''\ S=``3''\ path=``/article[1]/bdy[1]/sec[2]/ip1[1]''/$>$ \\
\ \ \ \ \ \ \ $<$path\ E=``3''\ S=``3''\ path=``/article[1]/bdy[1]/sec[2]/ss1[1]''/$>$ \\
\ \ \ \ \ \ \ $<$path E=``3'' S=``3'' path=``/article[1]/bdy[1]/sec[2]/ss1[1]/ip1[1]''/$>$ \\
\ \ \ \ \ \ \ $<$path E=``3'' S=``3'' path=``/article[1]/bdy[1]/sec[2]/ss1[1]/p[1]''/$>$ \\
\ \ \ \ \ \ \ $<$path E=``3'' S=``3'' path=``/article[1]/bdy[1]/sec[2]/ss1[2]''/$>$ \\
\ \ \ \ \ \ \ $<$path E=``3'' S=``3'' path=``/article[1]/bdy[1]/sec[2]/ss1[2]/ip1[1]''/$>$ \\
\ \ \ \ \ \ \ $<$path E=``3'' S=``3'' path=``/article[1]/bdy[1]/sec[2]/ss1[2]/p[1]''/$>$ \\
\ \ \ \ \ \ \ . . . . . . . \\
\ \ \ \ \ \ \ $<$path E=``3'' S=``3'' path=``/article[1]/bdy[1]/sec[4]''/$>$ \\
\ \ \ \ \ \ \ $<$path E=``0'' S=``0'' path=``/article[1]/bdy[1]/sec[4]/st[1]''/$>$ \\
\ \ \ \ \ \ \ $<$path E=``3'' S=``3'' path=``/article[1]/bdy[1]/sec[4]/ip1[1]''/$>$ \\
\ \ \ \ \ \ \ $<$path E=``3'' S=``3'' path=``/article[1]/bdy[1]/sec[4]/p[1]''/$>$ \\
\ \ \ \ \ \ \ $<$path E=``3'' S=``3'' path=``/article[1]/bdy[1]/sec[4]/p[2]''/$>$ \\
\ \ \ \ \ \ \ $<$path E=``3'' S=``3'' path=``/article[1]/bdy[1]/sec[4]/p[3]''/$>$ \\
\ \ \ \ \ \ \ . . . . . . . \\
\ \ \ \ \ \ \ $<$path E=``3'' S=``2'' path=``/article[1]/bm[1]''/$>$ \\
\ \ \ \ \ \ \ . . . . . . . \\
\ \ \ \ \ \ \ $<$path E=``3'' S=``2'' path=``/article[1]/bm[1]/app[1]''/$>$ \\
\ \ \ \ \ \ \ . . . . . . . \\
\ \ \ \ \ \ \ $<$path E=``3'' S=``2'' path=``/article[1]/bm[1]/app[1]/sec[1]''/$>$ \\
\ \ \ \ \ \ \ $<$path E=``3'' S=``2'' path=``/article[1]/bm[1]/app[1]/sec[1]/ip1[1]''/$>$ \\
$<$/file$>$ \\
\hline
\end{tabular}
\caption{INEX 2003 CO Relevance Assessment extract}
\label{fig-assessments}
\end{table}

\renewcommand\baselinestretch{1.5}

Table~\ref{fig-assessments} shows an extract of an INEX 2003 CO relevance assessment. The two INEX relevance dimensions, exhaustivity
\footnote{E represents the level of exhaustivity (with the values between 0-3)} and specificity
\footnote{S represents the level of specificity (with the values between 0-3)},
are applied to an article and elements within an article for the purpose of assessing their relevance to the CO topic. 

The focus of our analysis is on highly relevant elements,
which in our example are the elements with the value of 3
both for (E)xhaustiveness (the element covers ``most or all aspects of the topic'')
and (S)pecificity (``the topic ... is the only theme'' of the element)~\cite{INEX03RelAss}.
As shown in the example,
there are 15 such elements in this particular article, including the article itself.
These elements therefore constitute answer elements that are considered by the assessor as the
most preferable retrieval elements, even though 
there is a substantial amount of overlap between them. We can, however, identify
two complementary types of highly relevant elements: \emph{General} and \emph{Specific}.
Unlike the definitions for the S and E relevance dimensions that are provided by INEX, we define these types of highly relevant elements
as a result of our analysis as follows.

\begin{figure*}[tb]
\epsfxsize=12cm
\setlength{\epsfxsize}{\textwidth}
\centering\epsfbox{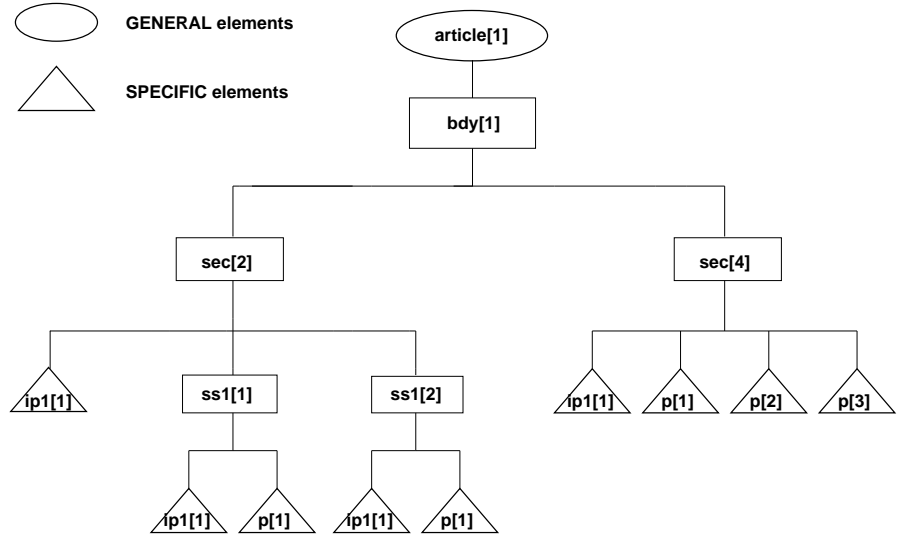}
\caption[GENERAL versus SPECIFIC elements: a tree-view example.]
{GENERAL versus SPECIFIC elements: a tree-view example.}
\label{fig-tree}
\end{figure*}

For a particular article in the collection, a \emph{General} element is
a least-specific highly relevant element containing other highly relevant elements.
In our example, \verb+article[1]+ is a General element,
since it is itself a highly relevant element, it contains all the other highly relevant elements,
and it is the least specific among them.
An article may contain several General elements if the article as a whole is not highly relevant.
Figure~\ref{fig-tree} shows a tree representation of all the highly relevant elements
shown in Table~\ref{fig-assessments}. The General element is the element shown in the ellipse.

For a particular article in the collection, a \emph{Specific} element is
a most-specific highly relevant element contained by other highly relevant elements.
In Figure~\ref{fig-tree}, the Specific elements
are the elements shown in triangle boxes.
Each of these elements is the most specific element among
all the other highly relevant elements that contain it.

When there is only one highly relevant element in an article,
that element is both a General and a Specific element.

INEX 2003 introduces 36 CO topics in total, with topic numbers between 91 and 126. We use the current version (2.5) of the INEX 2003 relevance assessments in our analysis. To date, no relevance assessments have been provided for topics 105, 106, 114 and 120. Moreover, topics 92, 100, 102, 118 and 121 do not contain highly relevant articles or elements within articles. Consequently, a total of 27 CO topics are used in our analysis.

\begin{figure*}[tb]
\epsfxsize=12cm
\centering\epsfbox{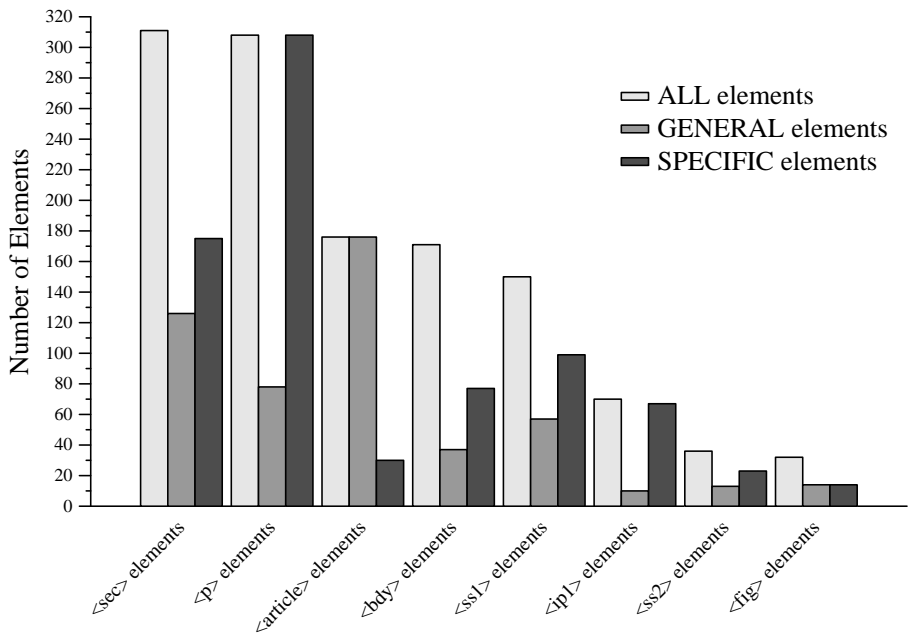}
\caption{INEX 2003 CO relevance assessments analysis: distribution of the most frequent highly relevant elements across all CO topics, for three distinct cases of relevance assessments.}
\label{fig-CO-number-elements}
\end{figure*}

Figure~\ref{fig-CO-number-elements} shows the distribution of the most frequent highly relevant elements within articles 
(including full articles) across all CO topics.
The figure shows three distinct cases when the relevance assessments consider all highly relevant elements (\emph{Original} relevance assessments),
General highly relevant elements only (\emph{General} relevance assessments), and Specific highly relevant elements only (\emph{Specific} relevance assessments), respectively.
The $x$-axis contains the names of the eight most frequent highly relevant elements in the case of Original relevance assessments.
The names of the elements correspond as follows:
\verb+<article>+ to a full article, \verb+<bdy>+ to article body,
\verb+<sec>+, \verb+<ss1>+ and \verb+<ss2>+ to section and subsection elements,
\verb+<p>+ and \verb+<ip1>+ to paragraph elements and \verb+<fig>+ to figure elements.
The $y$-axis contains the number of times each element occurs across the 27 CO topics.

Not surprisingly, in the case of Original relevance assessments,
the \verb+<sec>+ and \verb+<p>+ elements occur most frequently,
with 311 and 308 occurrences, respectively.
Perhaps surprisingly, the \verb+<article>+ and \verb+<bdy>+ elements are the next most frequent,
with 176 and 171 occurrences.
The latter suggests that in most cases when a \verb+<bdy>+ element is assessed as highly relevant,
the corresponding \verb+<article>+ element is also assessed as highly relevant.
This raises a question whether retrieving \verb+<bdy>+ elements make sense,
since for these cases the retrieval of \verb+<article>+ elements will equally satisfy the information need.
As shown in Figure~\ref{fig-CO-number-elements},
the next most frequent element is \verb+<ss1>+ with 150 occurrences, followed by
70 and below for the remaining elements.

In the case of General relevance assessments,
the situation changes in favour of least specific elements with \verb+<article>+ element being the most frequent.
The \verb+<sec>+, \verb+<p>+ and \verb+<ss1>+ elements follow next with
numbers significantly lower compared to the previous case (126, 78 and 57, respectively).
By looking at the number of \verb+<bdy>+ elements in this case,
we notice that there are at least 37 cases, across all CO topics,
when a \verb+<bdy>+ element is assessed highly relevant without a corresponding \verb+<article>+ element being assessed highly relevant too.

The last case shown on Figure~\ref{fig-CO-number-elements} is the case of Specific relevance assessments. In this case, as expected, the situation changes in favour of the most specific elements, with \verb+<p>+ element being most frequent.
The \verb+<sec>+ and \verb+<ss1>+ elements come next, followed by 77 occurrences of a \verb+<bdy>+ element. The latter observation suggests that there are situations where either \verb+<bdy>+ alone or a \verb+<bdy>+ and a corresponding \verb+<article>+ were the only elements assessed highly relevant. 

The above statistics provide an interesting insight of what might happen when the retrieval performance of an XML retrieval system 
is evaluated against three distinct cases of relevance assessments.
For instance, given the above knowledge of full articles being the most frequent highly relevant elements, 
one could expect that a full-text information retrieval system would be sufficient for satisfying users' information needs
in the case of General relevance assessments.
Conversely, in the case of Specific relevance assessments, an XML retrieval system capable of retrieving the most specific 
highly relevant elements within articles, such as a native XML database, should also be more than sufficient. 
The results of our experiments in Section 4 confirm the above expectations. Moreover, we show that our hybrid system, which combines the best
retrieval features from these two systems, can equally apply to either of the previously discussed assessment cases and is also capable of retrieving highly relevant document components.

\subsection{CO topic categories}

\begin{figure*}[tb]
\epsfxsize=9cm
\centering\epsfbox{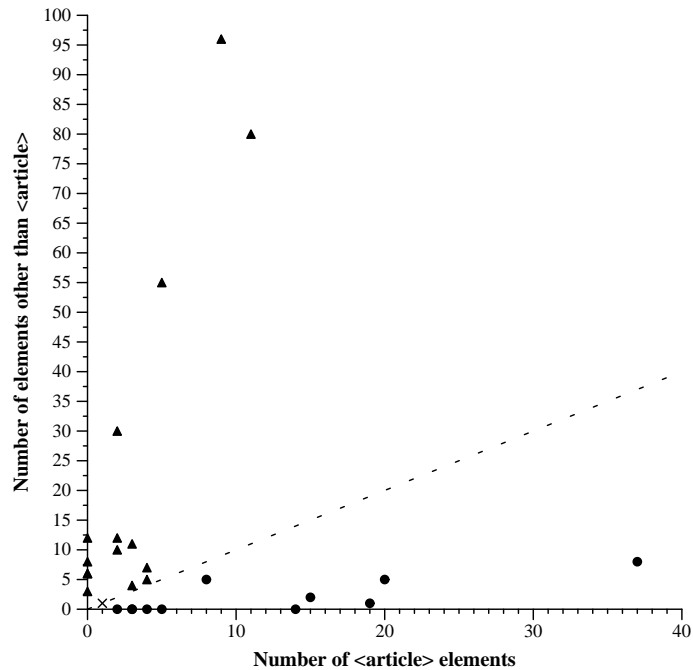}
\caption{INEX 2003 CO relevance assessments analysis: categories of CO topics when the relevance assessments consider General highly relevant elements only.}
\label{fig-CO-clustering}
\end{figure*}

In the following analysis we consider the case of General relevance assessments.
Our aim is to distinguish those CO retrieval topics that seek to mostly retrieve \verb+<article>+ elements
from those that mostly retrieve other, more specific elements within articles.
Consider the graph shown on Figure~\ref{fig-CO-clustering}.
The $x$-axis shows the number of General highly relevant \verb+<article>+ elements,
while the $y$-axis shows the number of General highly relevant elements other
than \verb+<article>+ contained by a CO topic.
A point on the graph therefore represents a particular CO topic.
For example, the CO topic depicted at coordinates (20,5) 
contains 20 highly relevant \verb+<article>+ elements
and 5 highly relevant elements other than articles.

We use this graph to identify two different categories of CO retrieval topics.

The first category of topics, shown as full circles on the graph and located below the dashed line, favour full article elements 
as highly relevant answers. There are 11 such topics (numbers 94, 95, 96, 97, 98, 107, 108, 110, 111, 115 and 122).
We refer to topics in this category as \emph{Broad} topics.

The second category of topics, shown as full triangles on the graph, favour elements other than full articles
as highly relevant answers.
We refer to topics in this category as \emph{Narrow} topics.
Note that the CO topic marked as \verb+x+ is neutral, since the numbers for both types of General highly relevant elements 
are the same.

The above topic categorisation cannot easily be derived in the other two assessment cases, that is, 
in the case of either the Original or the Specific relevance assessments. Indeed, due to 
the problem of overpopulated recall base the former case results in 
a situation where most CO topics contain far more specific highly relevant elements
than highly relevant articles.
The latter case is also not capable of identifying the above topic categories,
because of the very low number of highly relevant articles (\verb+30+ across all CO topics).
We believe that, given the various possible models of expected user behaviour, it is of great importance to 
recognise the different categories of CO retrieval topics that may exist for an XML document collection, 
since it is reasonable to expect varying performance of an XML retrieval system for either of the categories.
Indeed, the latter has been experimentally shown to be a valid assumption for a fragment-based XML retrieval system~\cite{HatanoINEX03}. 
We examine the above assumption in greater detail in Section 4.2, where we also compare the behaviour of different XML retrieval systems.

\section{XML Retrieval Approaches}

Most full-text information retrieval systems do not incorporate information about document structure.
Queries sent to such systems usually represent a bag of words, optionally including phrases or logical query operators.
The final answers are usually whole documents, presented in a descending order according to an estimate of their likelihood of relevance to
the information need expressed in the query. 

A native XML database provides strong support for storing and querying XML documents.
The information about the document structure is usually incorporated in the index structures, allowing
users to query both by document content and by document structure. Accordingly, elements belonging to particular articles
can easily be identified, either by the order they appear in the article or by certain keywords they contain.

To utilise the best retrieval features from both systems, we develop a \emph{hybrid} XML retrieval system
combining a full-text information retrieval system and a native XML database.

The following sections explore the three retrieval approaches in detail. We also identify and discuss some open issues that arise when a particular approach is applied to XML retrieval.

\subsection{A full-text information retrieval approach}

With a full-text information retrieval approach using Zettair,
the INEX XML document collection is first indexed by using an efficient inverted index structure.
Since Zettair stores the term postings file in a compressed form on disk, the size of the index takes roughly
26\% of the total collection size. Indexing the entire INEX collection on a system with Pentium 4 (2.66GHz processor) and 512MB RAM memory
running Mandrake Linux 9.1 takes around 70 seconds.

We use a  topic translation module to
automatically translate each INEX CO topic into a Zettair query.
For INEX CO topics, terms that
appear in the \verb+<Keywords>+ part are used to formulate the queries.
Up to 1500 \verb+<article>+ elements are then returned
in the descending order according to their estimated likelihood of relevance.
Accordingly, for a particular CO topic the above list of ranked articles represents Zettair's resulting answer list.

\begin{figure*}[tb]
\epsfxsize=9cm
\centering\epsfbox{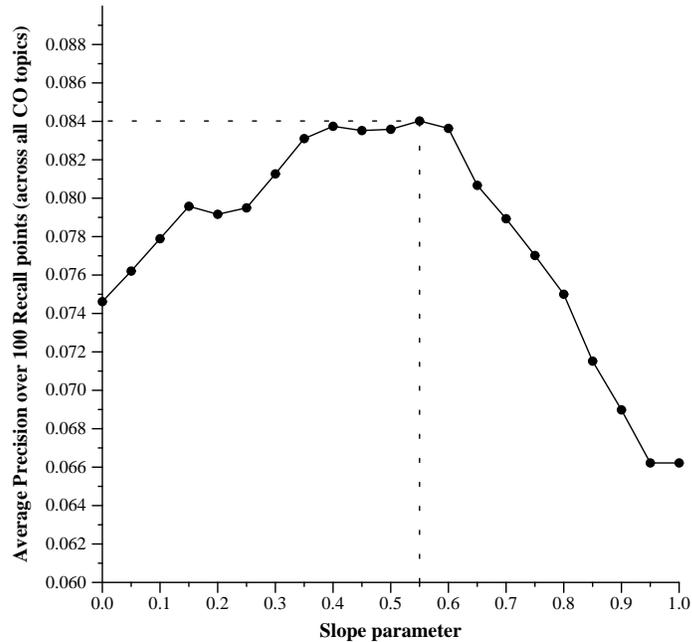}
\caption{Training pivoted cosine normalisation with Zettair on the INEX 2002 test collection.
The Mean Average Precision values are calculated using strict quantization function in inex\_eval.}
\label{Tuning-Zettair}
\end{figure*}

When retrieving documents of varying lengths,
the pivoted cosine document length normalisation scheme~\cite{Pivot} is shown to yield high retrieval effectiveness. The
pivoted cosine normalisation can be trained on the same collection by using a different set of retrieval topics
in order to identify the optimal \emph{slope} parameter.
Since the pivoted cosine similarity ranking formula is implemented in Zettair, we decided to train the pivoted cosine normalisation
on the INEX document collection by using the INEX 2002 CO retrieval topics. The INEX 2002 CO topics and their corresponding assessments
were consequently translated from INEX 2002 to INEX 2003 format.

The graph in Figure~\ref{Tuning-Zettair} shows the results of the pivoted cosine normalisation training process.
We use the strict quantization function in the \verb+inex_eval+ evaluation metric in our experiments~\cite{INEX02_overview}.
Numbers in the graph represent mean average precision values over 100 recall points across all INEX 2002 CO topics, when (up to) 1500 Zettair answers per CO topic are considered. The graph shows that a slope parameter with a value of 0.55 yields highest average precision value (0.0840). We therefore use the above value for the slope parameter in the Zettair pivoted cosine ranking formula for our experiments on the INEX 2003 CO topics.

\subsection{A native XML database approach}

Using a native XML database approach, the INEX XML document collection is first indexed by eXist.
Since the parsed document elements need also be stored in the index alongside the information for
all the element, attribute and word occurrences, the size of the index roughly doubles the total collection size.
Indexing the entire INEX collection on a system with Pentium 4 (2.6GHz processor) and 512MB RAM memory
running Mandrake Linux 9.1 takes around 2050 seconds.

We use a topic translation module to
automatically translate each INEX CO topic into two eXist queries: AND and OR.
The \verb+&=+ and \verb+|=+ eXist operators were used while formulating the above queries, respectively.
For INEX CO topics, the terms that
appear in the \verb+<Keywords>+ part are used to formulate eXist queries.
The AND answer list (that is, the answer list after using the AND query) constitutes
elements containing \emph{all} the query words or phrases; similarly, the OR answer list constitutes
elements containing \emph{any} of the query words or phrases.
However, because of the strict query conditions the AND answer list for most CO topics is likely to represent an empty list.
In fact, we have observed that all but 5 CO topics have empty AND answer lists.

The resulting answer list for a CO topic comprises elements from the AND answer list
followed by the elements from the OR answer list that do not belong to the AND answer list.
If an AND answer list is empty, the resulting answer list is the same as the OR answer list.
In either case it contains up to 1500 more specific elements within articles for each CO topic.
A sample of the resulting list of eXist matching elements is shown in Table~\ref{fig-eXistOR} (Section 3.4).

Despite numerous XML-specific retrieval advantages, we observe two serious issues with using eXist,
which are related to the XML retrieval process.

\begin{enumerate}

\item For a particular article in the answer list, eXist presents the final list of matching elements in article order,
according to the XQuery specification\footnote{http://www.w3.org/TR/xquery/\#N10895}. Moreover, if the answers are taken from more than
one article, eXist will order the articles by its own document identifier
that usually corresponds to the order in which
articles are stored in the database. We have noticed that some CO retrieval topics
produce a large list of matching elements, particularly in the case when OR queries are used (typically 10000 or more elements).
Each element in the answer list is therefore presented in article order,
with articles sorted by the document identifier.
This list is likely to constitute very many irrelevant elements as well as some relevant elements.
The relevant elements may belong to different articles, which could appear anywhere in the answer list.
Accordingly, finding articles estimated as likely to be \emph{relevant}
to the information need early in the retrieval process is not supported.

\item The matching elements that belong to a particular article in eXist's answer list represent
the most specific elements that satisfy the logical query conditions.
However, there is no additional information on whether a particular matching element is likely
to be more relevant than other matching elements in the list.
Accordingly, \emph{ranking} of the matching elements is not supported.
Moreover, there is no supporting information regarding the likelihood of relevance for the ancestor elements 
that contain the matching elements in the list.
eXist might then utilise the latter information in order to determine
the preferable level of \emph{granularity} for the final resulting elements.

\end{enumerate}

The following sections describe our approaches in addressing both these issues.

\subsection{A hybrid XML retrieval approach}

\begin{figure*}[tb]
\begin{center}
\epsfxsize=15cm
\setlength{\epsfxsize}{\textwidth}
\centering\epsfbox{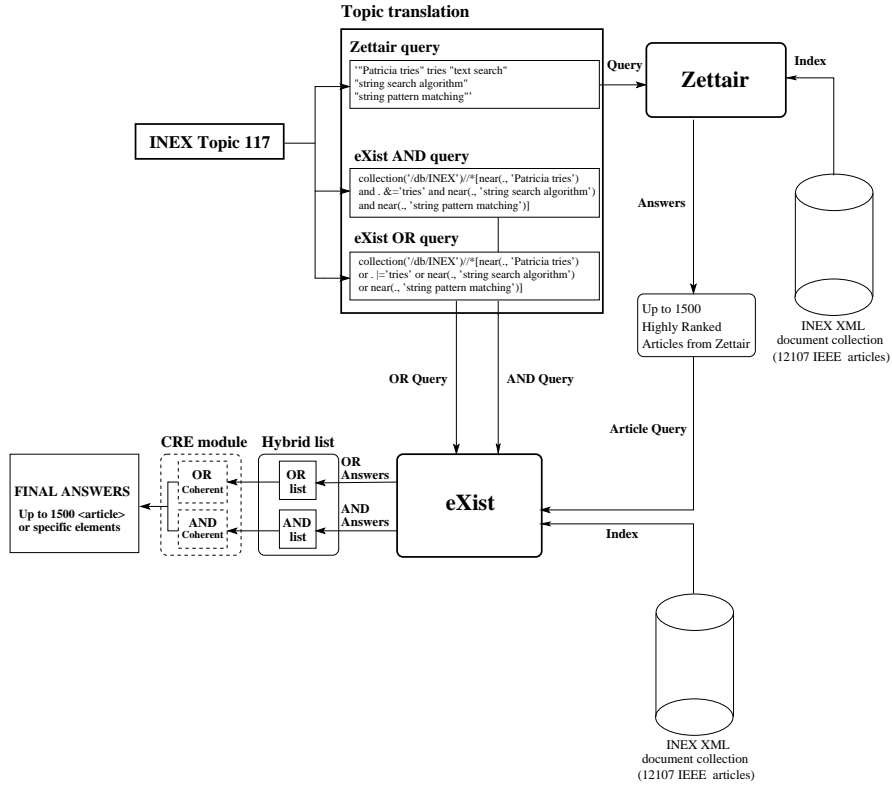}
\caption{A hybrid XML retrieval approach to INEX CO topics.}
\label{fig-hybrid}
\end{center}
\end{figure*}

Figure~\ref{fig-hybrid} shows that our hybrid approach is similar to the native XML database approach, except that
we first use Zettair to obtain up to 1500 articles likely to be considered relevant to the information need
expressed in the topic (that is, expressed in the translated Zettair query).
These articles are ranked in the descending order of their estimated likelihood of relevance.
For each article in the list, we use eXist to apply both queries, AND and OR, and produce matching elements in two corresponding answer lists.
Each answer list therefore comprises matching elements that belong to \emph{a particular article} in the list of articles returned by Zettair.
Accordingly, for each CO topic and a particular article,
the final answer list for an article comprises matching elements from the AND answer list
followed by the matching elements from the OR answer list that do not belong to the AND answer list.
The resulting answer list for each CO topic therefore comprises up to 1500 matching elements
taken from answer lists that belong to articles
that appear \emph{highest} in the ranked list of articles returned by Zettair.
The resulting answer list is shown as \verb+Hybrid list+ in Figure~\ref{fig-hybrid}.

The hybrid XML retrieval approach addresses the first retrieval problem
observed in a native XML database approach.
However, 
since our hybrid XML retrieval system uses eXist to produce its resulting answer list,
the second problem still remains open.
This raises an interesting question: is there a way of determining
which resulting elements are likely to represent preferable units of retrieval?

The following section describes one possible approach to providing an answer to this question.

\subsection{Ranking the native XML database output}

In order to effectively utilise the information contained in the resulting list of matching elements returned by eXist, 
we have developed a retrieval module that ranks the final answer elements. 
Since our module represents a post-processing module, it can equally be applied to both retrieval approaches (native XML database and hybrid). Moreover, it can easily be applied to any native XML database that preserves the list of its matching elements in article order.

\renewcommand\baselinestretch{}

\begin{table}[tb]
\begin{tabular}{l l}
\hline
Article & Answer element \\
\hline
ic/1999/w4095 & /article[1]/bdy[1]/sec[2]/ip1[1] \\
ic/1999/w4095 & /article[1]/bdy[1]/sec[2]/ss1[1]/ip1[1] \\
ic/1999/w4095 & /article[1]/bdy[1]/sec[2]/ss1[1]/p[1] \\
ic/1999/w4095 & /article[1]/bdy[1]/sec[2]/ss1[2]/p[1] \\
ic/1999/w4095 & /article[1]/bdy[1]/sec[2]/ss1[3]/ip1[1] \\
ic/1999/w4095 & /article[1]/bdy[1]/sec[4]/ip1[1] \\
ic/1999/w4095 & /article[1]/bdy[1]/sec[4]/p[1] \\
ic/1999/w4095 & /article[1]/bdy[1]/sec[4]/p[2] \\
ic/1999/w4095 & /article[1]/bdy[1]/sec[4]/p[3] \\
ic/1999/w4095 & /article[1]/bm[1]/app[1]/sec[1]/ip1[1] \\
ic/1999/w4095 & /article[1]/bm[1]/app[1]/sec[2]/p[1] \\
ic/1999/w4095 & /article[1]/bm[1]/app[1]/sec[2]/p[2] \\
\hline
\end{tabular}
\caption{eXist OR answer list example}
\label{fig-eXistOR}
\end{table}

\renewcommand\baselinestretch{1.5}

Within a particular article in the resulting answer list, we define a \emph{Coherent Retrieval Element (CRE)} as an element that contains \emph{at least} two matching elements, two other Coherent Retrieval Elements or a combination of a matching element and a Coherent Retrieval Element. In either case, the containing elements of a Coherent Retrieval Element should constitute either its \emph{different children} or each different child's \emph{descendants}.

There is one specific case, however.
If an answer list contains only one element, the CRE algorithm produces the same result: the matching element. This stems from the fact that there is no supporting information that will justify the choice for the ancestors of the matching element to be regarded as CREs.

Consider the eXist answer list shown in Table~\ref{fig-eXistOR}. The list shows eXist's matching elements
after the OR operator is used (which means each element in the list contains \emph{any} combination of keywords specified in the query).
Although the example shows a retrieval case when an OR list is used, the CRE algorithm equally applies in an AND list case.
Note that the matching elements in the answer list are presented in article order.

\begin{figure*}[tb]
\epsfxsize=12cm
\centering\epsfbox{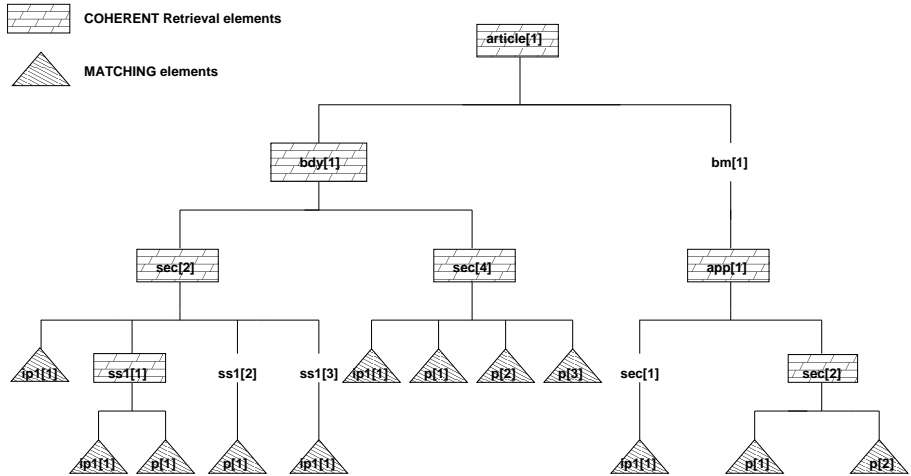}
\caption{Matching Elements versus Coherent Retrieval Elements: a tree-view example.}
\label{fig-coherent-tree}
\end{figure*}

Figure~\ref{fig-coherent-tree} shows a tree representation of the above list. The eXist matching elements are presented in triangle boxes, while the CREs are presented in square boxes. The figure also shows elements that are neither matching elements nor CREs.

By comparing the CREs shown in Figure~\ref{fig-coherent-tree} and the General and Specific (highly relevant) elements 
shown in Figure~\ref{fig-tree} in Section 2,
we observe that there are many cases where some matching elements represent Specific elements, while
there is one CRE representing a General element.
For example, the \verb+article[1]+ element represents both a General element and a CRE, and there are
eight matching elements that represent Specific elements.
Also, a reasonable assumption about the rest of the CRE's shown in square boxes in Figure~\ref{fig-coherent-tree} would be that they \emph{could} represent preferable units of retrieval for the CO retrieval topics. We validate this assumption in the next section.

Once the CREs have been identified, we still need to find a way to \emph{rank} and present them according to their \emph{estimated likelihood of relevance}.
We use a combination of the following XML-specific heuristics in order to determine the final rank of a CRE:

\begin{enumerate}
\item the number of times a CRE appears in each matching element's absolute path in the resulting answer list --
more matches (\verb+M+) or less matches (\verb+m+);
\item the length of the CRE's absolute path, taken from the root element --
longer path (\verb+P+) or shorter path (\verb+p+); and
\item the ordering of the XPath sequence in the CRE's absolute path --
nearer to beginning (\verb+B+) or nearer to end (\verb+E+).
\end{enumerate}

\renewcommand\baselinestretch{}

\begin{table}
\begin{tabular}{lcccccccc}
\hline
& \verb+Mp+ & \verb+MP+ & \verb+mp+ & \verb+mP+ & \verb+Pm+ & \verb+PM+ & \verb+pm+ & \verb+pM+ \\
\hline
\hline
\verb+B+ & 0.1203 & 0.1199 & 0.0792 & 0.0785 & 0.0792 & 0.0819 & 0.1015 & 0.1136 \\
\verb+E+ & \bf{0.1209} & 0.1206 & 0.0817 & 0.0792 & 0.0801 & 0.0826 & 0.1019 & 0.1143 \\
\hline
\end{tabular}
\caption{Mean Average Precision values for the hybrid-CRE retrieval system, as different CRE heuristic combinations apply. 
The values are generated using strict quantization function in inex\_eval in the case of Original CO relevance assessments.}
\label{CRE-heuristics}
\end{table}

\renewcommand\baselinestretch{1.5}

There are eight distinct cases of heuristic combinations that can be explored in order to determine 
the final rank of a CRE, provided the ordering of the heuristics is preserved as above. However, we also expect that 
a reordering of the above heuristics could determine different CRE's ranks. 
We therefore analyse all possible CRE heuristic combinations
(16 in total, since we regard the third heuristic as complementary to the other two and therefore always apply at the end).
Table~\ref{CRE-heuristics} shows this analysis and the impact that different heuristic combinations have on the overall retrieval effectiveness. 
The values in the table represent mean average precision values over 100 recall points across all CO topics and are generated using
strict quantization function in the \verb+inex_eval+ evaluation metric~\cite{INEX02_overview}.
We use the case of Original relevance assessments. The CRE module is applied on the 
resulting answer list of our hybrid XML retrieval system, and all the retrieved CREs (per article) are included in the final answer list.

As shown in Table~\ref{CRE-heuristics}, 
the best effectiveness is produced by using the
\verb+Mp+ or the \verb+pM+ heuristic combination, with the former producing higher scores.
For both cases, using the \verb+E+ heuristic (as complementary 
to the above heuristic combinations) produces better scores than using the \verb+B+ heuristic. Thus, the best heuristic combination is \verb+MpE+.

\renewcommand\baselinestretch{}

\begin{table}[tb]
\begin{tabular}{l l c c r}
\hline
Article & Answer element & Matches & Length & Sequence \\
\hline
ic/1999/w4095 & /article[1] & 12 & 1 & 1 \\
ic/1999/w4095 & /article[1]/bdy[1] & 9 & 2 & 11 \\
ic/1999/w4095 & /article[1]/bdy[1]/sec[2] & 5 & 3 & 112 \\
ic/1999/w4095 & /article[1]/bdy[1]/sec[4] & 4 & 3 & 114 \\
ic/1999/w4095 & /article[1]/bm[1]/app[1] & 3 & 3 & 111 \\
ic/1999/w4095 & /article[1]/bdy[1]/sec[2]/ss1[1] & 2 & 4 & 1121 \\
ic/1999/w4095 & /article[1]/bm[1]/app[1]/sec[2] & 2 & 4 & 1112 \\
\hline
\end{tabular}
\caption{Ranked list of Coherent Retrieval Elements}
\label{fig-CoherentOR}
\end{table}

\renewcommand\baselinestretch{1.5}

This heuristic combination can be interpreted as follows: first the CREs are sorted in a descending order according to the number of times each 
CRE appears in the resulting list of matching elements (the more often it appears, the better). 
Next, if two Coherent Retrieval Elements appear the same number of times in the answer list, the one with the shorter length will be ranked higher.
If, however, they have the same length, the ordering sequence where they appear in the article will determine their final ranks.
For example, if it so happens that \verb+article[1]/bdy[1]/sec[2]+ and \verb+article[1]/bdy[1]/sec[4]+ appear 
the same number of times in the answer list and have the same length, then when using the above heuristic combination, the 
latter element will be ranked higher than the former element.

After applying our retrieval module on the answer list shown in Table~\ref{fig-eXistOR}, we produce our ranked list of Coherent Retrieval Elements
as shown in Table~\ref{fig-CoherentOR}.

This table also shows that our retrieval module can also determine the level of \emph{granularity}
of the CREs in the ranked list.
With the current implementation, less specific and more general CREs
are preferred over more specific and less general CREs.
However, we observe that different heuristic combinations may be more suitable 
for a different choice of evaluation metric (such as \verb+inex_eval_ng(s)+ or \verb+inex_eval_ng(o)+),
in which case our retrieval module could easily be switched to produce
more specific and least general CREs early in the ranking 
(such as using the \verb+PME+ heuristic combination, which as we show in Section 4.3
appears to be more suitable for both the above metrics). 

The ranked list of CREs is also shown as \verb+CRE module+ in Figure~\ref{fig-hybrid}.

\section{Experiments and Results}

In order to determine the most effective content-only XML retrieval system,
we investigate the following systems:

\begin{itemize}

\item Zettair, using a full-text information retrieval approach;
\item eXist, using a native XML database approach;
\item Hybrid, using a hybrid approach to XML retrieval;
\item eXist-CRE, using a native XML database approach with our Coherent Retrieval Element (CRE) module applied on the resulting answer list; and
\item Hybrid-CRE, using a hybrid XML retrieval approach with the CRE module applied on the resulting answer list.

\end{itemize}

For each of the above, the resulting answer list for a particular CO topic comprises
up to 1500 articles or elements within articles.
A retrieval run for each retrieval system therefore constitutes resulting answer lists for all CO topics.
In order to evaluate the effectiveness of each system,
for each CO topic the average precision value (over 100 recall points) is first calculated.
These values are then averaged over all CO topics, and the mean average precision value for a particular run is produced.
The strict quantization function in a respective 
evaluation metric is used in order to calculate the mean average precision values.

\subsection{Comparison of retrieval approaches}

\renewcommand\baselinestretch{}

\begin{table*}[tp]
\begin{center}
\begin{tabular}{c c c c c c}
\hline
 \bf{Maximum} & \multicolumn{1}{c}{\bf{eXist}} & \multicolumn{1}{c}{\bf{eXist-CRE}} & \multicolumn{1}{c}{\bf{Hybrid}} & \multicolumn{1}{c}{\bf{Hybrid-CRE}} & \multicolumn{1}{c}{\bf{Zettair}} \\
\cline{2-2} \cline{3-3} \cline{4-4} \cline{5-5} \cline{6-6}
\bf{retrieved} & $matching$ & $Coherent$ & $matching$ & $Coherent$ & $articles$ \\
\bf{elements} & elements & elements & elements & elements & only \\
\bf{(per article)} & only & only & only & only & \\
\hline \hline
1 & 0.0013 & 0.0055 & 0.0063 & 0.0512 & \bf{0.0520} \\
2 & 0.0017 & 0.0063 & 0.0064 & 0.0926 &  \\
3 & 0.0018 & 0.0070 & 0.0072 & 0.1078 & \\
4 & 0.0018 & 0.0079 & 0.0090 & 0.1122 & \\
5 & 0.0018 & 0.0081 & 0.0123 & 0.1153 & \\
6 & 0.0018 & 0.0084 & 0.0130 & 0.1206 & \\
7 & 0.0018 & 0.0087 & 0.0146 & 0.1221 & \\
8 & 0.0018 & 0.0086 & 0.0158 & 0.1215 & \\
9 & 0.0018 & 0.0087 & 0.0175 & 0.1231 & \\
10 & 0.0019 & 0.0088 & 0.0190 & \bf{0.1256} & \\
11 & 0.0020 & 0.0087 & 0.0208 & 0.1248 & \\
12 & 0.0021 & 0.0087 & 0.0227 & 0.1238 & \\
13 & 0.0021 & 0.0087 & 0.0233 & 0.1229 & \\
14 & 0.0022 & 0.0088 & 0.0243 & 0.1223 & \\
15 & 0.0023 & 0.0088 & 0.0258 & 0.1214 & \\
16 & 0.0023 & 0.0089 & 0.0264 & 0.1220 & \\
17 & 0.0025 & 0.0090 & 0.0272 & 0.1225 & \\
18 & 0.0025 & 0.0090 & 0.0276 & 0.1236 & \\
19 & 0.0026 & 0.0090 & 0.0280 & 0.1234 & \\
20 & 0.0026 & 0.0090 & 0.0285 & 0.1231 & \\
\hline
\verb+all+ (up to 1500) & \bf{0.0028} & \bf{0.0091} & \bf{0.0322} & 0.1209 & \\
\hline \hline
\end{tabular}
\end{center}
\caption{Mean Average Precision values for different XML retrieval systems, as the number of retrieved elements per article increase.
The values are generated using strict quantization function in inex\_eval in the case of Original CO relevance assessments.}
\label{LuXist-table}
\end{table*}

\renewcommand\baselinestretch{1.5}

Table~\ref{LuXist-table} shows the retrieval runs for each XML retrieval system, 
when using the \verb+inex_eval+ evaluation metric
and the case of Original relevance assessments.
The \verb+MpE+ heuristic combination is implemented in the CRE retrieval module. 
The best runs for each system are highlighted in bold.
As shown in the table, we determine the optimal number of retrieved elements (per article) for each retrieval system except Zettair, since
its only unit of retrieval is a full article.
Although for plain eXist and the plain Hybrid system retrieving more elements increases the effectiveness,
the case when all (up to 1500) elements are retrieved performs best.
This is also the case for eXist-CRE.
Contrary to the above, in the case when the hybrid system implements the CRE module (Hybrid-CRE), the optimal number
of retrieved CREs (per article) is 10. 
However, in the next sections we show that, for a fixed heuristic combination (\verb+MpE+ in this case) and 
different retrieval scenarios,
the value for the optimal number of retrieved CREs can not be easily determined.
 
The table also shows that the Hybrid-CRE system is by far the most effective.
This supports our previous observation that
for a particular CO topic and a particular article,
the list of CREs produced by our CRE module indeed represents a list of preferable units of retrieval and,  
in most of the cases, the list contains the full article element itself.
On the other hand, plain eXist is the least effective system, while
our plain Hybrid system,
which uses eXist as a central retrieval module, is roughly 11 times
more effective than plain eXist. This shows that a native XML database can effectively utilise
information about articles \emph{estimated as likely to be relevant} to a particular CO topic.
The results also show that when our CRE module is applied, eXist's effectiveness is more than three times greater than that of the plain eXist.
Compared to eXist-CRE, our Hybrid-CRE system 
also improves its retrieval effectiveness by roughly 14 times.
These are all significant improvements for the retrieval effectiveness of a native XML database system.

For Zettair, the above retrieval results clearly identify the importance of
having a full-text information retrieval system in the XML retrieval task.
If we compare the best result of our plain Hybrid system,
we notice that Zettair still performs better.
This is not surprising, since our analysis of the INEX 2003 CO topics relevance assessments shows
that the full article element represents the third most frequent highly relevant element
among all the highly relevant elements in the INEX 2003 XML test collection.
It then follows that, for the CO topics, Zettair is indeed capable of retrieving highly relevant articles,
while our plain Hybrid system retrieves other elements (in the case of its best run up to 1500), 
which are more specific and not necessarily highly relevant.
However, if the CRE module is applied to the Hybrid system,
we observe an effectiveness improvement of roughly 2.5 times more than that of Zettair, and roughly 4 times 
more than that of the plain Hybrid system.
The latter results show that without the CRE module, the native XML database in the plain Hybrid system is not
capable of identifying the \emph{coherent} highly relevant elements for the CO topics.

\begin{figure*}[tb]
\epsfxsize=14cm
\centering\epsfbox{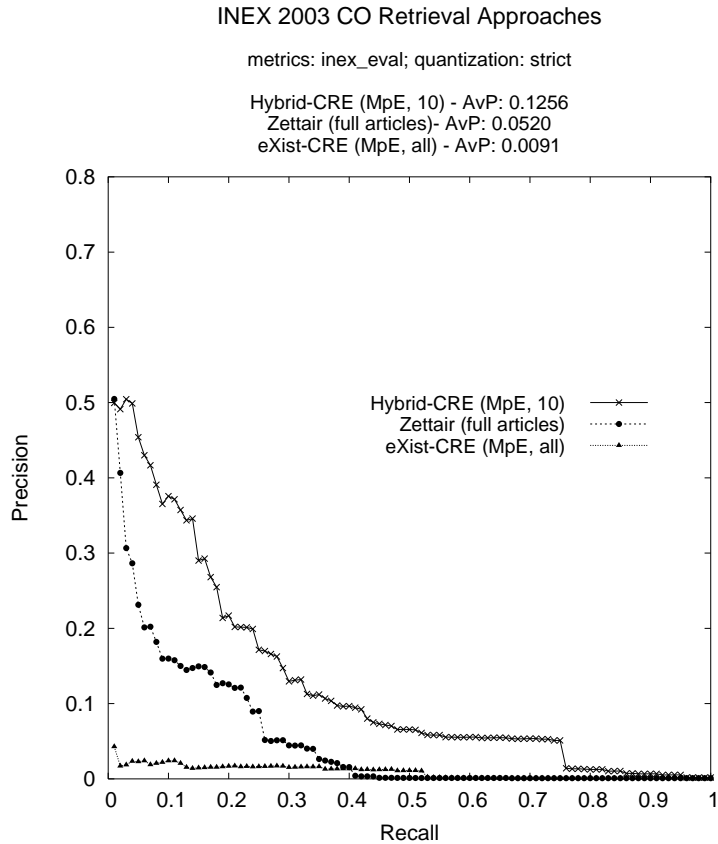}
\caption{Evaluation of the XML retrieval systems using strict quantization function in inex\_eval 
and the case of Original CO relevance assessments.}
\label{fig-inex-eval}
\end{figure*}

Figure~\ref{fig-inex-eval}
shows a detailed summary of the best evaluation results for the XML retrieval systems (for each of the three approaches).
As observed previously in Table~\ref{LuXist-table}, the hybrid-CRE system
performs best, followed by Zettair, and the eXist-CRE system is worst.

\subsection{Analysis based on different retrieval scenarios}

\renewcommand\baselinestretch{}

\begin{table}
\begin{tabular}{lcccc}
\hline
 & & \multicolumn{3}{c}{\bf{inex\_eval}} \\
\cline{3-5}
 & & \multicolumn{3}{c}{\bf{Original assessments}} \\
\cline{3-5}
\bf{XML retrieval approach} & \bf{\verb+n+} & $~All$ topics & $~Broad$ topics & $~Narrow$ topics \\
\hline
\hline
\verb+Zettair+ & \verb+1+ &  0.0520 & 0.0884 & 0.0270 \\
\hline
 & \verb+1+ & 0.0013 & 0.0012 & 0.0014 \\
\verb+eXist+ & \verb+10+ & 0.0019 & 0.0013 & 0.0024 \\
 & \verb+all+ & 0.0028 & 0.0013 & 0.0038 \\
\hline
 & \verb+1+ & 0.0055 & 0.0110 & 0.0018 \\
\verb+eXist-CRE (MpE)+ & \verb+10+ & 0.0088 & 0.0171 & 0.0030 \\
 & \verb+all+ & 0.0091 & 0.0170 & 0.0036 \\
\hline
 & \verb+1+ & 0.0063 & 0.0061 & 0.0017 \\
\verb+Hybrid+ & \verb+10+ & 0.0190 & 0.0252 & 0.0147 \\
 & \verb+all+ & 0.0322 & 0.0339 & 0.0311 \\
\hline
 & \verb+1+ & 0.0512 & 0.0830 & 0.0294 \\
\verb+Hybrid-CRE (MpE)+ & \verb+10+ & \bf{0.1256} & \bf{0.1492} & 0.1094 \\
 & \verb+all+ & 0.1209 & 0.1348 & \bf{0.1114} \\
\hline \hline
\end{tabular}
\caption{Mean Average Precision values for the XML retrieval systems over different CO topic categories
 with three different numbers of retrieved elements (per article).
The values are generated using strict quantization function in inex\_eval in the case of Original relevance assessments.}
\label{Original-categories}
\end{table}

\renewcommand\baselinestretch{1.5}

In the following analysis, 
we evaluate the effectiveness of each XML retrieval system
for three different cases of relevance assessments: 
 Original, General and Specific. 
For each relevance assessment case, 
the performance of each system is compared using all CO topics (\emph{All}),
or using either of the two 
categories of CO topics: \emph{Broad} and \emph{Narrow} (these categories were previously identified in Section 2.3).
For each XML retrieval system (except Zettair),
 we additionally investigate three different numbers of retrieved elements per article (\verb+n+): 
 \verb+1+, \verb+10+ and \verb+all+. 
 We choose these values because 
 \verb+n=10+ works well with the previously determined CRE heuristic combination, while the other two
 represent the lowest and the highest possible values of \verb+n+.

\renewcommand\baselinestretch{}

\begin{table}
\begin{tabular}{lcccc}
\hline
 & & \multicolumn{3}{c}{\bf{inex\_eval}} \\
\cline{3-5}
 & & \multicolumn{3}{c}{\bf{General assessments}} \\
\cline{3-5}
\bf{XML retrieval approach} & \bf{\verb+n+} & $~All$ topics & $~Broad$ topics & $~Narrow$ topics \\
\hline
\hline
\verb+Zettair+ & \verb+1+ & \bf{0.1694} & \bf{0.3178} & 0.0674 \\
\hline
 & \verb+1+ & 0.0003 & 0.0003 & 0.0003  \\
\verb+eXist+ & \verb+10+ & 0.0003 & 0.0003 & 0.0003  \\
 & \verb+all+ & 0.0003 & 0.0003 & 0.0003 \\
\hline
 & \verb+1+ & 0.0085 & 0.0186 & 0.0015 \\
\verb+eXist-CRE (MpE)+ & \verb+10+ & 0.0046 & 0.0091 & 0.0015 \\
 & \verb+all+ & 0.0045 & 0.0091 & 0.0014 \\
\hline
 & \verb+1+ & 0.0003 & 0.0003 & 0.0004 \\
\verb+Hybrid+ & \verb+10+ & 0.0007 & 0.0003 & 0.0010 \\
 & \verb+all+ & 0.0027 & 0.0003 & 0.0043 \\
\hline
 & \verb+1+ & 0.1438 & 0.2614 & 0.0629 \\
\verb+Hybrid-CRE (MpE)+ & \verb+10+ & 0.0786 & 0.0940 & \bf{0.0680} \\
 & \verb+all+ & 0.0746 & 0.0852 & 0.0673 \\
\hline \hline
\end{tabular}
\caption{Mean Average Precision values for the XML retrieval systems over different CO topic categories
 with three different numbers of retrieved elements (per article).
The values are generated using strict quantization function in inex\_eval in the case of General relevance assessments.}
\label{General-categories}
\end{table}

\renewcommand\baselinestretch{1.5}

Table~\ref{Original-categories} shows the evaluation results for each system in the case of Original relevance assessments. 
Overall, the Hybrid-CRE system performs best. However, the optimal number of retrieved 
elements changes from 10 (when the system is evaluated against \emph{All} and \emph{Broad} topics)
  to \verb+all+ (when the system is evaluated against the \emph{Narrow} topics). 
Not surprisingly, when evaluated against the \emph{Narrow} topics the plain Hybrid system performs better than Zettair.
For the \emph{Broad} topics, the systems implementing the CRE module preserve their
optimal number of retrieved elements at 10, while the other systems still need to retrieve the maximum number of elements to achieve 
their best performances.

\renewcommand\baselinestretch{}

\begin{table}
\begin{tabular}{lcccc}
\hline
 & & \multicolumn{3}{c}{\bf{inex\_eval}} \\
\cline{3-5}
 & & \multicolumn{3}{c}{\bf{Specific assessments}} \\
\cline{3-5}
\bf{XML retrieval approach} & \bf{\verb+n+} & $~All$ topics & $~Broad$ topics & $~Narrow$ topics \\
\hline
\hline
\verb+Zettair+ & \verb+1+ & 0.0021 & 0.0023 & 0.0020 \\
\hline
 & \verb+1+ & 0.0012 & 0.0006 & 0.0015  \\
\verb+eXist+ & \verb+10+ & 0.0020 & 0.0007 & 0.0029  \\
 & \verb+all+ & 0.0035 & 0.0008 & 0.0054 \\
\hline
 & \verb+1+ & 0.0013 & 0.0017 & 0.0011 \\
\verb+eXist-CRE (MpE)+ & \verb+10+ & 0.0034 & 0.0063 & 0.0013 \\
 & \verb+all+ & 0.0033 & 0.0063 & 0.0013 \\
\hline
 & \verb+1+ & 0.0063 & 0.0122 & 0.0023 \\
\verb+Hybrid+ & \verb+10+ & 0.0347 & 0.0524 & 0.0225 \\
 & \verb+all+ & \bf{0.0566} & \bf{0.0689} & \bf{0.0481} \\
\hline
 & \verb+1+ & 0.0038 & 0.0020 & 0.0050 \\
\verb+Hybrid-CRE (MpE)+ & \verb+10+ & 0.0262 & 0.0311 & 0.0228 \\
 & \verb+all+ & 0.0243 & 0.0248 & 0.0240 \\
\hline \hline
\end{tabular}
\caption{Mean Average Precision values for the XML retrieval approaches over different CO topic categories
 with three different numbers of retrieved elements (per article).
The values are generated using strict quantization function in inex\_eval in the case of Specific relevance assessments.}
\label{Specific-categories}
\end{table}

\renewcommand\baselinestretch{1.5}

Table~\ref{General-categories} shows the evaluation results for each system in the case of General relevance assessments.
We observe two things in both cases of \emph{All} and \emph{Broad} topics: 
first, the Zettair system performs best,
significantly outperforming the other systems particularly for the latter topic category; and second, 
the optimal number of retrieved elements for systems implementing the CRE module is 1.
However, the situation changes for the \emph{Narrow} topics, where 
the hybrid-CRE run (with 10 retrieved elements per article)
performs slightly better than Zettair. The latter observation thus confirms our previous claim
about an expected varying performance of XML retrieval systems when evaluated against categories of CO topics.
Indeed, even the retrieval parameter (\verb+n+), implemented in our CRE module, should have a different setting 
when the system is evaluated against the \emph{Narrow} CO topic category.

In the case of Specific relevance assessments, 
there is no clear distinction of the (possible) categories of CO topics. 
Indeed, Table~\ref{Specific-categories} shows that in this case the plain Hybrid system performs best 
(although the absolute performance scores are rather poor),
irrespective to the knowledge of the existing categories of CO topics. 
We also observe that, for both \emph{Broad} and \emph{Narrow} topics,
the systems perform best when retrieving the maximum number of elements.

\subsection{Analysis based on different evaluation metrics}

The aim of this section is to provide an insight into
the behaviour for each of the XML retrieval systems,
when their performance is evaluated using 
the following evaluation metrics: \verb+inex_eval_ng(s)+ and \verb+inex_eval_ng(o)+~\cite{inex_eval_ng}. 
These metrics include information about the sizes of the retrieved elements;
the latter taking into account the (possible) extent of overlapping between elements in the resulting answer list.

\renewcommand\baselinestretch{}

\begin{table}
\begin{tabular}{lccc}
\hline
&  & \multicolumn{1}{c}{\bf{inex\_eval\_ng(s)}} & \multicolumn{1}{c}{\bf{inex\_eval\_ng(o)}} \\
\cline{3-3} \cline{4-4} 
 & & \multicolumn{1}{c}{\bf{Original}} & \multicolumn{1}{c}{\bf{Original}} \\
\cline{3-3} \cline{4-4} 
\bf{XML retrieval approach} & \bf{\verb+n+} & $All$ topics & $All$ topics\\
\hline
\hline
\verb+Zettair+ & \verb+1+ & 0.0933 & 0.1252 \\
\hline
& \verb+1+ & 0.0089 & 0.0090 \\
\verb+eXist+ & \verb+10+ & 0.0132 & 0.0138 \\
& \verb+all+ & 0.0150 & 0.0161 \\
\hline
& \verb+1+ & 0.0251 & 0.0286 \\
\verb+eXist-CRE (MpE)+ & \verb+10+ & 0.0337 & 0.0286 \\
& \verb+all+ & 0.0342 & 0.0286 \\
\hline
& \verb+1+ & 0.0168 & 0.0182 \\
\verb+Hybrid+ & \verb+10+ & 0.0978 & 0.1087 \\
& \verb+all+ & 0.1057 & 0.1141 \\
\hline
& \verb+1+ & 0.1044 & 0.1344 \\
\verb+Hybrid-CRE (MpE)+ & \verb+10+ & 0.2086 & 0.1423 \\
& \verb+all+ & \bf{0.2156} & \bf{0.1473} \\
\hline \hline
\end{tabular}
\caption{Mean Average Precision values for the XML retrieval approaches using different INEX evaluation metrics
 with three different numbers of retrieved elements (per article).
The values are generated using strict quantization function in the case of Original relevance assessments and across all CO topics.}
\label{strict-metrics}
\end{table}

\renewcommand\baselinestretch{1.5}

Table~\ref{strict-metrics} shows the performance scores for the XML retrieval systems,
in the case of Original relevance assessments and
across all CO topics, when the effectiveness is evaluated using these evaluation metrics.
The CRE module uses the \verb+MpE+ heuristic combination, a combination which produced the best performance scores 
for the \verb+inex_eval+ evaluation metric.
For each metric and each system, 
we also investigate the optimal number of retrieved elements per article: \verb+1+, \verb+10+ or \verb+all+.

For both evaluation metrics, our Hybrid-CRE system again performs best,
and for each system the optimal number of retrieved elements is the maximum number (\verb+all+),
 irrespective to whether the system incorporates the CRE module. 
However, the ordering of the systems in terms of how they perform differs for each metric. 
For example, when the \verb+inex_eval_ng(s)+ metric is used, the best run for the plain Hybrid system outperforms the Zettair run,
which is not the case when the \verb+inex_eval_ng(o)+ metric applies. This is rather strange, since these systems do not produce 
overlapping elements in their resulting answer lists. 

Furthermore, despite the many overlapping elements in the resulting list
 of the Hybrid-CRE system, it is still ranked higher than all the other systems. 
This raises the question whether the CRE module should utilise another, more suitable heuristic combination.
Indeed, these metrics utilise a fundamentally different 
interpretation of the concept of relevance than that of
 the \verb+inex_eval+ metric; the relevance concept is referred to as an ``ideal concept space''~\cite{inex_eval_ng}. 
 We therefore experimented with different possible heuristic combinations and found that, for these metrics,
 the best heuristic combination is \verb+PME+ (Section 3.4 explores all the possible heuristic combinations). 
 When implemented in the CRE module, our Hybrid-CRE system produces the following mean average precision values: 0.2201 for \verb+ng(s)+ and
 0.1640 for \verb+ng(o)+, yielding an 11\% relative performance improvement over the previous 
 heuristic combination when the \verb+ng(o)+ metric applies. This is perhaps not surprising, 
 since the \verb+PME+ heuristic retrieves more specific CREs early in the ranking, therefore reducing the penalising effect of the latter metric.
 Consequently, tuning an XML retrieval system with appropriate values for the retrieval parameters 
 (such as the \verb+PME+ choice for the heuristic combination, in this case) depends on the choice of evaluation metric. 
 
\section{Related work}

\subsection{Comparison to other INEX systems}

The participants in INEX 2003 used various approaches to XML retrieval. These approaches were
classified in two categories: \emph{model-oriented} and \emph{system-oriented}~\cite{INEX03_overview}.
Our group followed a system-oriented approach by using the initial hybrid XML retrieval system, where
we investigated various extraction strategies with eXist~\cite{LuXistINEX03}.
These strategies resulted in rather poor system effectiveness for CO topics, where Zettair
performed better than our initial hybrid system.
Wilkinson also shows that simply
extracting components from likely relevant documents leads to poor system
performance~\cite{RossW}. However, the hybrid system with CRE module (which we developed after INEX 2003)
more than doubles the retrieval effectiveness of a full-text information retrieval system.
We show elsewhere that the above approach also produces further performance improvements for the CAS topics~\cite{TDM04}.

At INEX 2002 the CSIRO participating group proposed a similar approach to XML retrieval~\cite{Padre}.
Queries are sent to PADRE, the core of the CSIRO Panoptic Enterprise
Search Engine\footnote{http://www.panopticsearch.com}. Unlike Zettair, where the primary unit of retrieval is a
full article, PADRE combines full-text and metadata indexing and retrieval and is capable of indexing particular 
elements within articles, such as \verb+<author>+, \verb+<sec>+ and \verb+<p>+.
However, in contrast to our CRE retrieval module,
the above approach ignores the higher structural elements that contain the indexed element.

The following is a brief summary of some related XML retrieval approaches presented at INEX 2003. 
XXL~\cite{XXLINEX03} implements a simple probabilistic ranked XML retrieval and additionally incorporates ontological knowledge 
for both the element names and the element content. The Otago system~\cite{Otago} implements  
a proprietary corpus tree structure which is first used to rank the likely relevant documents and then, 
using the notion of ``coverage'', extracts likely relevant elements 
from those documents. To correctly deal with ranked retrieval on the document component level,
the IBM system~\cite{IBMINEX03} implements an extension of the classical vector space model by merging ranked answers
coming out from different, previously determined component indexes. The ILLC-Amsterdam group uses a retrieval system 
that implements a multinomial language model for determining the preferable units of retrieval and their final result scores~\cite{BorkurINEX03}.
The EXTIRP system~\cite{EXTIRP} first splits the XML documents into a set of ``minimal XML fragments'' and then ranks them using 
a similarity measure. The rank values are accordingly propagated to the ancestor elements and the final ranked list of resulting 
fragments is generated. The system also includes a query expansion module. The SearX engine~\cite{SearX} is a commercial, out-of-the-box XML search 
engine which is based on the probabilistic retrieval principle. To handle different document fragments
sharing a common semantics, as well as incorporate weightings into the query language, SearX utilises the concept of \emph{roles}.
This concept is shown to work well in a publishing environment where large and structured document collections are used.

When compared to their initial INEX 2003 performance scores for different choice of evaluation metrics~\cite{INEX03_overview}, 
we have found that our newly developed Hybrid-CRE system is the most effective among all the above XML retrieval systems.
The best performance scores outlined in Table~\ref{LuXist-table} and Table~\ref{strict-metrics} were used for the Hybrid-CRE system 
with reference to the above comparison.

\subsection{Preferable units of retrieval}
Some related research has also been carried out in an effort to determine the preferable units of XML retrieval for the CO topics.
The focused structured document retrieval aims to identify and retrieve
``best entry points'' that are considered as document components from which users can easily browse and access highly relevant
information~\cite{BEP}. They investigate the best retrieval strategies derived from user studies that
effectively identify the best entry points.
However, the best entry points do not necessarily represent highly relevant document components,
since they primarily capture the relationship between the retrieved document components.
In XIRQL~\cite{XIRQL} the preferable units of retrieval, referred to as ``index objects'', are specified by domain analysts manually analysing
the XML document schema. Contrary to the manual approach, Hatano et al. propose a statistical analysis that determines
``Coherent Partial Documents'' from the INEX XML document collection without need of a document schema~\cite{HatanoINEX02}.
However, when compared to our CRE retrieval module, which for a particular CO topic utilises the structural information from
the matching elements and dynamically determines the Coherent Retrieval Elements,
Hatano et al.'s analysis requires the entire XML document collection to be processed.

\subsection{INEX 2003 CO topic categories}
Hatano et al. have also undertaken an analysis of INEX 2003 CO relevance assessments, where they also focus 
on highly exhaustive and highly specific elements~\cite{HatanoINEX03}. Their analysis is primarily 
based upon the statistics gathered for the number of the answer XML document components and the average number of words per document component
for each CO topic. On the basis of their analysis, they have also identified two different categories of CO topics: 
\verb+SCO+, topics for ``searching specific XML fragments'', and \verb+ACO+, topics for ``searching aggregated XML fragments''. 
The SCO and ACO topic categories therefore correspond to the Broad and the Narrow topic categories we have previously identified, 
although our analysis is based on somewhat different foundation. 
More precisely, roughly 64\% of the total number of topics that appear in our Broad category
also appear in the ACO category, and roughly 67\% of the total number of topics that appear in our Narrow category
also appear in the SCO category.
They have also identified the ``nested relationships'' among the XML components in the recall base, an issue 
previously identified as the overlap problem in XML retrieval evaluation. They argue that depending on its retrieval purpose,
an XML retrieval system should be evaluated against those topics that fulfill that purpose.
We have also evaluated Zettair and eXist, two systems with different retrieval purposes, 
against different categories of CO topics. However, unlike their ``XML fragment retrieval system'', we have 
proposed a hybrid XML retrieval system that is still capable of 
identifying and retrieving highly relevant document components, irrespective to the existing CO topic categories.

\subsection{Full-text search and ranking}
Most native XML databases implement efficient storage and querying mechanisms over XML documents and (usually) retrieve document components
that strictly match the logical query conditions. Most have also implemented a support for XQuery -- the standard XML query language~\cite{XQuery}.
Recently, the World Wide Web Consortium has acknowledged the need for incorporating full-text search and ranking capabilities 
into XQuery~\cite{FT-Use, FT-Req}.
The TeXQuery language~\cite{TeXQueryLang}, implemented in the Quark\footnote{http://www.cs.cornell.edu/database/quark/} system, and the TIX bulk-algebra~\cite{TIX}, implemented in Timber\footnote{http://www.eecs.umich.edu/db/timber/}, represent initiatives aiming at integrating 
information retrieval techniques into a standard native XML database query evaluation engine.
TeXQuery specifies a bi-directional mapping between the XQuery data model and its formal data model. 
The two additional TeXQuery expressions enable users to express the full-text search queries and 
to additionally rank the resulting answers.
The TIX algebra, on the other hand, is based on the notion of a ``scored tree'', which represents a rooted tree where each node 
incorporates at least two attributes, indicating the name and the relevance score of the node. 
Additional operators manipulate the scored data trees in an information retrieval fashion;
by enabling retrieval of elements 
after satisfying the requirements for the score or the rank position, and
by specifying the way of selecting the most appropriate among all the likely relevant elements 
(much like the algorithm implemented in our CRE module).
The TeXQuery language and the TIX algebra thus attempt to bridge the gap between 
the information retrieval and the strict database approaches to XML retrieval. 
The two retrieval concepts have also been implemented in fully functional demo prototypes~\cite{TeXQueryDemo, TIXDemo}.

The main contribution of the XXL search engine~\cite{XXL} is an ontology-enabled search, 
which can span across XML document collections each conforming to a different underlying schema. However, 
it is unclear whether XXL is capable of determining the appropriate retrieval granularity of 
the likely relevant document components, particularly during the query evaluation phase.

In an effort to improve information retrieval of whole documents in a structured document environment, 
Trotman develops a corpus-tree data structure that 
utilises the underlying structural information during retrieval~\cite{Trotman}. 
The similarity ranking measures are also slightly adapted in order to support the 
structured document search process. 
However, additional 
index structures and further adaptation of the implemented similarity measures may be needed in order to retrieve and rank document
components instead of whole documents.

XSearch~\cite{XSearch} retrieves answers
consisting of semantically related nodes. However, it cannot be used (without modification) in a straightforward fashion 
with the INEX XML document collection, since the 
structural document schema for the INEX collection does not incorporate element semantics.
XRank~\cite{XRank}, on the other hand, follows the structured retrieval principle and aims at returning 
the most specific ancestor elements, which comprise matching elements containing any of the query keywords, 
irrespective to the underlying document schema.
However, it is primarily tuned for a hyperlinked XML environment and is thus unknown whether it is capable of producing highly effective scores
in a non-hyperlinked environment such as the INEX XML document collection.

\section{Conclusion and future work}

This paper investigates the implications that arise when 
the performances of three systems, a full-text information retrieval system, a native XML database, and a hybrid system, 
are evaluated against different XML retrieval scenarios.

The different cases of relevance assessments, which were identified as a result of our INEX 2003 relevance assessments analysis, 
can approximate different (possible) models of user behaviour;
we have shown that the parameters and the performance of an XML retrieval system may vary widely, depending on which particular 
retrieval model is used. Moreover, the knowledge of the existing topic categories can, in some assessment cases, 
also influence the choice of the optimal retrieval parameter(s). 
We have also shown that tuning an XML retrieval system depends on the choice of evaluation metric.
In fact, since different evaluation metrics usually tend to model different retrieval behaviours, it is not feasible, nor indeed possible,
to find the best combination of retrieval parameters that could work equally well with every metric. 
Some efforts are being made, however, in the direction of 
unifying existing INEX metrics into
a robust evaluation metric which aims at modeling
all the possible situations of expected retrieval behaviour~\cite{Overlap}.

Since the values for the mean average precision for the XML retrieval systems 
(including ours) are still very low compared to the values for systems retrieving whole documents, 
it is our hope that the work presented in this paper 
will lead to a better understanding of issues concerning the relevance assessments and the choices for
the optimal system parameters in different XML retrieval scenarios.

In the future,
we aim to implement a relevance ranking scheme in eXist and investigate whether or not it would be more efficient as well as more effective solution.

\acknowledgements
We thank Wolfgang Meier for his assistance with using eXist, and
Nick Lester, Falk Scholer and other members of the Search Engine Group at RMIT
for their support with using Zettair.
Further, we thank Saied Tahaghoghi and the anonymous reviewers for providing useful comments
on earlier drafts of this paper.

\end{article}

\end{document}